\documentclass{sn-jnl}%
\usepackage{graphicx}  
\usepackage{dcolumn}  
\usepackage{bm}    
\usepackage{amsmath}
\usepackage{physics}
\usepackage{hhline}
\usepackage{multirow}
\usepackage{booktabs}
\usepackage{hyperref}   
\usepackage{amstext}
\usepackage{orcidlink}

\begin{document}

\title{Evaluating alternative spin scenarios of $P_{c\bar{c}}(4440)$ and $P_{c\bar{c}}(4457)$ using heavy quark symmetries}
\author{\fnm{Duygu} \sur{ Y\i{}ld\i{}r\i{}m}\orcidlink{0000-0001-5499-9727}}\email{yildirimyilmaz@amasya.edu.tr}

\affil{\orgdiv{Physics Department}, \orgname{Faculty of Sciences and Arts}, \orgaddress{\street{ Amasya University}, \city{Amasya}, \postcode{05200},  \country{Turkey}}}


\abstract{
Heavy quark symmetries are useful for predicting the existence of heavy states, their masses, and spin states. Despite numerous studies on the $P_{c\bar{c}}(4440)$ and $P_{c\bar{c}}(4457)$ heavy states, their spin states have not been previously determined. In this study, heavy symmetries are applied to predict the spin states. If the $P_{c\bar{c}}(4440)$ and $P_{c\bar{c}}(4457)$ states are considered $\bar{D}^*\Sigma_c$ molecules, they can be classified as heavy partners. This classification may help clarify their potential connections with heavy antiquark-diquark symmetry partners. By utilizing these  alternative spin assignments  and the concept of heavy antiquark-diquark symmetry, it may be possible to estimate $\Xi_{cc}^{(*)}\Sigma_c^{(*)}$ states, and ultimately, their spin states, which have not been elucidated in experiments. In addition to these symmetries, the relationship between $P_{c\bar{c}}$ and $P_{c\bar{c}s}$ pentaquarks can be constructed which supports the prediction of possible $P_{c\bar{c}s}$ states. The predicted masses of the $P_{c\bar{c}s}(4338)$ and $P_{c\bar{c}s}(4459)$ states align with several studies, allowing us to eliminate a specific spin state. One spin state appears to be favored, suggesting  that $P_{c\bar{c}}(4440)$ has $J^P=\frac{1}{2}$ and $P_{c\bar{c}}(4457)$ has $J^P=\frac{3}{2}$.}


\maketitle 

\section{Introduction}

Interest in exotics increased after the discovery of $\chi_{c1}(3872)$~\cite{PhysRevLett.91.262001}, which was called exotic because it did not easily fit into the traditional quark model~\cite{GELLMANN1964214}. As a result, some alternative definitions have been proposed, such as tetraquark~\cite{Faccini:2012pj} and hadrocharmonium~\cite{PhysRevD.72.114013,Swanson:2004pp}. The state has primarily been treated as a hadronic molecule because of its proximity to the $D^{*}\bar{D}^*$ threshold and its decay to $J/ \Psi 2\pi(3 \pi)$ further strengthens the interpretation of a molecule~\cite{Braaten:2003he,Belle:2011vlx,Gamermann:2009uq}. In addition to $\chi_{c1}(3872)$, many exotic states have been observed and are considered hadronic molecules, including $P_{c\bar{c}}(4312)$, $P_{c\bar{c}}(4380)$, $P_{c\bar{c}}(4440)$, $P_{c\bar{c}}(4457)$, $P_{c\bar{c}s}(4338)$, $P_{c\bar{c}s}(4459)$, $T_{c\bar{c}}(4020)$~\cite{PhysRevLett.111.242001}, $T_{b\bar{b}1}(10610)$, $T_{b\bar{b}1}(10650)$~\cite{Yildirim:2023znd, PhysRevD.91.072003, PhysRevLett.108.122001}, and others~\cite{RevModPhys.90.015004}.

In the last decade, a few hidden-charm pentaquarks, namely $P_{c\bar{c}}(4312)$, $P_{c\bar{c}}(4380)$, $P_{c\bar{c}}(4440)$, and $P_{c\bar{c}}(4457)$, have been discovered. They are observed in the $J/ \Psi p$ invariant mass distribution of decay $\Lambda_b \rightarrow J / \Psi p K^- $~\cite{LHCb:2015yax,LHCb:2019kea}. All these states include at least five quarks $uud\bar{c}\bar{c}$. Therefore, they were initially considered pentaquarks ~\cite{PhysRevD.101.014002,PhysRevD.100.016014,Wang:2019got,Mutuk:2019snd,PhysRevD.100.054002}. However, because of the closeness of the $P_{c\bar{c}}(4312)$, $P_{c\bar{c}}(4440)$, and $P_{c\bar{c}}(4457)$ states (simply referred to as $P_{c\bar{c}}$ states to avoid excessive repetition) to the $\bar{D}\Sigma_c$ and $\bar{D}^*\Sigma_c $ thresholds, with small decay widths~\cite{PhysRevD.110.030001}:

\begin{eqnarray} \label{eq:1}
 m_{P_{c\bar{c}}(4312)}&=&4311.9 \hspace{0.2 cm} \text{MeV},  \hspace {0.5 cm} \Gamma_{P_{c\bar{c}}(4312)}=9.8 \hspace{0.2 cm} \text{MeV}    \, ,   \nonumber \\
 m_{P_{c\bar{c}}(4440)}&=&4440.3 \hspace{0.2 cm} \text{MeV}  , \hspace{0.5  cm}  \Gamma_{P_{c\bar{c}}(4440)}=20.6 \hspace{0.2 cm} \text{MeV}    \, ,  \\
 m_{P_{c\bar{c}}(4457)}&=&4457.3 \hspace{0.2  cm}  \text{MeV}  , \hspace{0.5  cm}  \Gamma_{P_{c\bar{c}}(4457)}=6.4  \hspace{0.2 cm} \text{MeV}  \, .\nonumber  
\end{eqnarray}
Although they are generally considered  to be hadronic molecules~\cite{Azizi:2020ogm, PhysRevLett.124.072001, PhysRevD.100.014022, PhysRevD.100.014021, Chen:2019bip}, other suggestions exist for these states, such as hadrocharmonium \cite{Eides:2019tgv} and virtual states \cite{PhysRevLett.123.092001}.

Furthermore, the spin states of these hadronic molecules have not been clarified, but knowing the spin state is crucial for distinguishing them. The spin states of $P_{c\bar{c}}(4440)$ and $P_{c\bar{c}}(4457)$ cannot be determined with the available data, they are predicted to be $J^P=\frac{1}{2}^-$ or $J^P=\frac{3}{2}^-$. Several studies have attempted to determine their spin states using various approaches. Some studies have shown that the spin of $P_{c\bar{c}}(4440)$ is $J^P=\frac{1}{2}^-$ and the spin of $P_{c\bar{c}}(4457)$ is $J^P=\frac{3}{2}^-$~\cite{PhysRevD.100.014022,Chen:2019bip,PhysRevLett.122.242001,Chen:2019asm,PhysRevD.108.L031503,  ZHANG2023981,Wang:2022oof,PhysRevD.103.034003}, while others have shown the opposite designation~\cite{Yamaguchi:2019seo, PhysRevD.103.054004,PhysRevD.102.114020,Uchino:2015uha,Karliner_2015}.  It was also suggested that  $P_{c\bar{c}}(4440)$ and $P_{c\bar{c}}(4457)$ are $J^P=\frac{3}{2}^-$ or $P_{c\bar{c}}(4440)$ is $\Sigma \bar{D}$ with $J^P=\frac{1}{2}^-$ and $P_{c\bar{c}}(4457)$ is $\Sigma^* \bar{D}^*$ with $J^P=\frac{5}{2}^-$~\cite{Chen:2019bip}. Additionally, the selection of quantum numbers changes by including one pion contribution ~\cite{PhysRevD.100.094028}. With only contact interactions, $P_{c\bar{c}}(4440)$ is $J^P=\frac{1}{2}^-$ and $P_{c\bar{c}}(4457) $ is $J^P=\frac{3}{2}^-$,  and one pion exchange contribution changes the situation in reverse. Because one pion exchange is attractive(repulsive) for the $\frac{3}{2}^-(\frac{1}{2}^-)$ channel, the preferred spin state of $P_{c\bar{c}}(4440)$ is $\frac{1}{2}^-$ and that of $P_{c\bar{c}}(4457)$ is $\frac{3}{2}^-$. There are even proposals with positive parity definitions for $P_{c\bar{c}}(4440)$ and $P_{c\bar{c}}(4457)$~\cite{Chen:2015moa, Xiang:2017byz}. Overall, the spin states of the $P_{c\bar{c}}$ states have not been definitively determined.

The symmetries of effective field theories derived from quantum chromodynamics (QCD) provide a practical approach to understanding exotic states. These symmetries, particularly in the heavy quark limit, are crucial for predicting the possible heavy states. In this study, two heavy quark symmetries are applied. The first applied heavy quark spin symmetry (HQSS) aids in identifying and characterizing molecular states in the heavy quark sector. HQSS implementation significantly reduces the number of independent interaction terms required in the model, thereby simplifying the theory and making it more predictive~\cite{ISGUR1989113}. The second applied heavy antiquark-diquark symmetry (HADS) is beneficial for establishing a relationship between the properties of hadrons with one heavy antiquark and two heavy quarks~\cite{SAVAGE1990177}. The observation of three of the six possible heavy quark spin partners in $D^{(*)}\bar{D}^{(*)}$ demonstrates the successful application of heavy hadron symmetries.

The emergence of various exotic states, such as $P_{c\bar{c}}$, $P_{c\bar{c}s}$, and $T_{b\bar{b}1}$, suggests the possibility of specific heavy hadron-hadron systems that exhibit similar interactions and binding mechanisms. Although $\Sigma_c$ and $\Xi_c$ belong to different $SU(3)_f$ multiplets, the $P_{c\bar{c}}$ and $P_{c\bar{c}s}$ states can be related through generalized flavor-spin symmetry~\cite{Chen:2021spf}. This symmetry allows for an investigation of the similarities and differences between the $P_{c\bar{c}}$ and $P_{c\bar{c}s}$ states~\cite{Chen:2022wkh}, which have similar binding energies and dependences on the effective potential. Furthermore, the $P_{c\bar{c}}$ and $P_{c\bar{c}s}$ states are considered $\bar{D}^{(*)}\Sigma_c^{(*)}$ and $\bar{D}^{(*)} \Xi^{(*)} $ molecules respectively, and they are rooted in HQSS restrictions. The total effective potentials obtained from the light degrees of freedom are given along with their potentials in Table~\ref{tab:table1}. From this perspective, the quantum numbers are assigned as  $P_{c\bar{c}}(4440)$ $\frac{1}{2}$ and $P_{c\bar{c}}(4457)$ $\frac{3}{2}$. Additionally, a recent study proposed that if $P_{c\bar{c}s}(4338)$ is regarded as a $\frac{1}{2}^-$ $ \bar{D} \Xi_c $ molecule, the potentials of $\frac{1}{2}^-$ $\bar{D} \Xi_c$, $\frac{1}{2}^-$ $ \bar{D}^* \Xi_c$, and $\frac{3}{2}^-$ $ \bar{D}^*\Xi_c$ are identical within the heavy quark symmetry framework~\cite{Burns:2022uha}.
\begin{table}[htbp]
	\caption{\label{tab:table1}  Heavy quark spin partners of $\bar{D}^{(*)}\Sigma_c^{(*)}$ and $\bar{D}^{(*)} \Xi^{(*)}_c$ states associated with generalized flavor-spin symmetry, along with their shared potentials. The threshold values are given in units of MeV.}
	\begin{tabular}{lcccccc} 
		\toprule
		Molecule & $J^P$ & Threshold  &   Molecule & $J^P$ & Threshold & V  \\ \midrule \\[-2.0ex]   
		$\bar{D} \Sigma_c$  & $\frac{1}{2}^-$ &  $4322$  &  $\bar{D} \Xi_c $ &   $\frac{1}{2}^-$  &   $4337$ & $C_a$ \\[1.0ex]   
		$\bar{D} \Sigma_c^*$ & $\frac{3}{2}^-$  &  $4386$ & $\bar{D}^* \Xi_c $ & $\frac{1}{2}^- /  \frac{3}{2}^-$ &  $4478$  & $C_a$   \\[1.0ex] 
		$\bar{D}^* \Sigma_c$ & $\frac{1}{2}^-$  & $4463$ & $\bar{D}^* \Xi^{\prime}_c $ &$\frac{1}{2}^-$ & $4587$ &  $C_a - \frac{4}{3}C_b$ \\[1.0ex] 
		$\bar{D}^* \Sigma_c$ & $\frac{3}{2}^-$   & $4463$  & $\bar{D}^* \Xi^{\prime}_c $ &$\frac{3}{2}^-$ & $4587$ & $C_a + \frac{2}{3}C_b$  \\[1.0ex]  
		$\bar{D}^*\Sigma_c^*$ & $\frac{1}{2}^-$  &  $4527$   & $\bar{D}^* \Xi^*_c $ &$\frac{1}{2}^-$ & $4655$ & $C_a - \frac{5}{3}C_b$   \\[1.0ex] 
		$\bar{D}^*\Sigma_c^*$ & $\frac{3}{2}^-$   &  $4527$  & $\bar{D}^* \Xi^*_c $ &$\frac{3}{2}^-$ & $4655$ & $C_a - \frac{2}{3}C_b$    \\[1.0ex] 
		$\bar{D}^*\Sigma_c^*$ & $\frac{5}{2}^-$  &  $4527$  & $\bar{D}^* \Xi^*_c $ &$\frac{5}{2}^-$ & $4655$ & $C_a+C_b$  \\ 
		\bottomrule
	\end{tabular}
\end{table}

Resonances are usually analyzed using the Breit-Wigner parameterization, although other methods have been developed, such as Flatt\'e  parameterization and effective range expansion. Until recently, many researchers believed that the Breit-Wigner approach most effectively explained the experimental data. However, some researchers have expressed concerns regarding the compatibility of unitarity and analyticity when evaluating resonances near the threshold. It has also been suggested that this approach may distort the resonance shape; thus, Breit-Wigner can lead to ambiguities when applied to near-threshold states. Notably, extracting resonance values is only appropriate using this method if the resonance is above the threshold of interest~\cite{Adachi:2011mks,PhysRevLett.108.122001}, but reports show that $\chi_{c1}(3872)$~\cite{Braaten:2007dw}, $T_{b\bar{b} 1}^{(\prime)}$~\cite{Cleven:2011gp,Cleven:2013rkf}, and $P_{c\bar{c}s}(4338)$~\cite{Meng:2022wgl} may not be suitable using the Breit-Wigner distribution, mainly because the parameterization introduces  difficulties when studying resonances close to the threshold value. To overcome these difficulties, an alternative method called Sill~\cite{Giacosa:2021mbz} has recently been proposed. Sill is superior in some cases, particularly when dealing with $XYZ$ states. 

The present study builds on earlier work~\cite{PhysRevD.98.114030}. In particular, we examine the heavy quark symmetry partners of $\Xi_{cc}^{(*)}\Sigma_c^{(*)}$ states, assuming that $P_c(4450)$ is a $\bar{D}^*\Sigma_c$ molecule, within the HQSS and HADS frameworks. It is assumed that they are formed as hadronic molecules, which interact only in the contact range. Because Refs.~\cite{Valderrama:2012jv,Lu:2017dvm} studies indicated that pion exchange contributions are perturbative in the charm sector.

With increasing data in 2019, the $P_c(4450)$ state was split into two states: $P_{c\bar{c}}(4440)$ and $P_{c\bar{c}}(4457)$; however their  spin states are unclear. This study aims to determine the spin states of $P_{c\bar{c}}(4440)$ and $P_{c\bar{c}}(4457)$, which have been the focus of several previous studies following their discovery. We assume that the $P_{c\bar{c}}(4312)$, $P_{c\bar{c}}(4440)$, and $P_{c\bar{c}}(4457)$ states are s-wave antimeson baryon hadronic molecules, and their heavy quarks make them amenable to HQSS and HADS analyses. The HADS partners are analyzed because they are expected to provide valuable insights into the spin of the $P_{c\bar{c}}$ states. Specifically, it is expected that if $\bar{D}^{(*)}\Sigma_c^{(*)}$ forms a bound state, then the $\Xi_{cc}^{(*)}\Sigma_c^{(*)}$ states should also be bound. Therefore, a thorough analysis of the $\Xi_{cc}^{(*)}\Sigma_c^{(*)}$ states may provide information regarding the spin of the $P_{c\bar{c}}$ states. We also derive supporting information from the relationship between the  $P_{c\bar{c}}$ and $P_{c\bar{c}s}$ states to determine their spin states. As needed, using the Sill parameterization, which differs from the Breit-Wigner approach through a redefinition of resonance parametrization, this study explores, the potential spin states of $P_{c\bar{c}}(4440)$ and $P_{c\bar{c}}(4457)$ based on all available information and assumptions.

Herein, the possible spin states of $P_{c\bar{c}}$ have been  more precisely identified owing to the contribution of heavy quark symmetries and a new resonance parameter. The remainder of this paper is organized as follows. We introduce our theoretical framework in Section~\ref{sec:2}, present and discuss the corresponding numerical results in Section~\ref{sec:3}, and summarize the study in Section~\ref{sec:4}. 

 \section{Theoretical Framework}\label{sec:2}
 
 We employ effective field theory to investigate heavy states. It is assumed that they are formed as hadronic molecules, which interact only in the contact range. Previous studies indicated that pion exchange contributions are perturbative in the charm sector \cite{PhysRevD.98.114030,Yan2022}.
 
HQSS imposes constraints on the structure of the interactions between heavy hadrons, stating that interactions should not depend on the spin of the heavy quark. This symmetry leads to the conclusion that pseudoscalar $D$ and vector $D^*$ heavy mesons are degenerate states. Consequently, the states composed of $D$ and $D^*$  can be placed into one $2\times 2$ nonrelativistic superfield matrix and expressed as follows~\cite{FALK1992119<}:
\begin{equation} \label{eq:2}
H_c=\frac{1}{\sqrt{2}}[D+\vec{D}^*\cdot \vec{\sigma}] \, ,
\end{equation}  
where $\vec{\sigma}$ denotes the Pauli matrices. The $H_c$ superfield respects heavy quark rotations. 
 
For the heavy baryon under HQSS, the superfield consisting of the total spin $S=\frac{1}{2}$ ground state $\Sigma$  and the  $S=\frac{3}{2}$ excited state $\Sigma^*$ is given as a $2\times 3$ matrix~\cite{CHO1994683}:
\begin{equation}\label{eq:3}
    \vec{S}_c=\frac{1}{\sqrt{3}} \vec{\sigma}\Sigma_c +\vec{\Sigma}^*_c \, .
\end{equation}
The fact that $\Sigma^*$ satisfies the condition $\vec{\sigma}\cdot \vec{\Sigma}_c^*=0$ ensures that $ \vec{\Sigma}_c^*$ is a spin $\frac{3}{2}$ state.  With the $H_c$ and $\vec{S}_c$ superfields,  the Lagrangian containing the contact range interaction without derivatives can be written as~\cite{PhysRevD.98.114030}
\begin{equation} \label{eq:4}
\mathcal{L}= C_a \vec{S}^{\dagger}\cdot \vec{S} Tr[\bar{H}^{\dagger}\bar{H}]+C_b\sum_{i=1}^{3}\vec{S}^{\dagger}\cdot (J_i \vec{S}) Tr [\bar{H}^{\dagger} \sigma_i \bar{H}]  \, ,
\end{equation}
where $C_a$ and $C_b$ are low-energy coupling constants.  $J_i$ with $i=1,2,$ and $3$ are the spin-1 angular momentum matrices. This Lagrangian leads to seven contact $\bar{D}^{(*)}\Sigma^{(*)}$ heavy molecules. The obtained molecule states and their contact potentials are listed in Table \ref{tab:table1}.  For simplicity, the isospin is ignored in the computations.

HADS, which provides a significant advancement in our understanding of particle physics, suggests that a heavy diquark $(QQ)$ behaves like a heavy antiquark $(\bar{Q})$. This approach not only provides a deeper insight into the behavior of heavy states but also offers a new perspective on the structure of heavy baryons. A baryon's heavy diquark component has a length scale of $1/(m_Qv)$ and can be considered point-like if the quarks are heavy enough, given that its length scale is smaller than the typical QCD length scale of $ 1/\Lambda_{QCD}$. The light-quark cloud around a heavy diquark in a heavy baryon is similar to that around a heavy antiquark in a heavy antimeson. This situation leads to HADS, which has profound implications for the hadron structure and properties. Hence, HADS relates hadrons between content with the same light quarks but different numbers of heavy quarks. For instance, the properties of heavy antimesons, specifically  $(\bar{D},\bar{D}^*)$, are connected to doubly heavy baryons, $ ( \Xi_{cc},  \Xi_{cc}^* )$, through HADS. The \(\bar{D}^{(*)}\Sigma_c^{(*)}\) states can be viewed as extensions of the \(\Xi_{cc}^{(*)}\Sigma_c^{(*)}\) states through the \(D^{(*)}\) meson. As a result, in the heavy quark limit, the contact potential of the \(\bar{D}^{(*)}\Sigma_c^{(*)}\) state is associated with the \(\Xi_{cc}^{(*)}\Sigma_c^{(*)}\) state. This symmetry enables assumptions about the masses of these states.

For the $\Xi_{cc}$ and $\Xi_{cc}^*$ states, referring to the spin $\frac{1}{2}$ and spin $\frac{3}{2}$ states of doubly heavy baryons, the definition of the  nonrelativistic superfield is given as~\cite{PhysRevD.73.054003}
\begin{equation} \label{eq:5}
    \vec{T}_{cc}=\frac{1}{\sqrt{3}} \vec{\sigma} \Xi_{cc} + \vec{\Xi}_{cc}^* \, .
\end{equation}
Although $\vec{T}_{cc}$  is analogous to $\vec{S}_c$, they differ in terms of spin content. The $\vec{S}_c$' light spin is $1$ and  the heavy spin is $\frac{1}{2}$,  whereas the light spin of $\vec{T}_{cc}$ is $\frac{1}{2}$ and the heavy spin is $1$. The Lagrangian in Eq.\ref{eq:4} can be adopted by applying HADS \cite{SAVAGE1990177},
\begin{equation}\label{eq:6}
 Tr \left[ \bar{H}^{\dagger}\bar{H} \right] \rightarrow  \vec{T}^{\dagger}\cdot \vec{T}  \hspace{0.5 cm} Tr \left[ \bar{H}^{\dagger} \sigma_i \bar{H}\right] \rightarrow  \vec{T}^{\dagger}\cdot ( \sigma_i \vec{T}) \, . 
\end{equation}
Using these transformations, as previously described in Ref.~\cite{PhysRevD.73.054003, PhysRevD.98.014014}, the contact range Lagrangian for the interaction of a heavy baryon and a doubly heavy baryon is given as~\cite{PhysRevD.98.114030}
\begin{equation} \label{eq:7}
    \mathcal{L}=C_a \vec{S}^{\dagger}\cdot \vec{S} \vec{T}^{\dagger}\cdot \vec{T}  + C_b\sum_{i=1}^{3}\vec{S}^{\dagger}\cdot (J_i \vec{S}) \vec{T}^{\dagger} \cdot (\sigma_i \vec{T}) \, .
\end{equation}
The states corresponding to the Lagrangian and its corresponding contact range potentials are given in Table~\ref{tab:table3}.

Employing HADS, it is possible to predict the mass relationship between these states~\cite{PhysRevD.73.054003},
\begin{equation}  \label{eq:8}
    m_{\Xi_{cc}^*}- m_{\Xi_{cc}}=\frac{3}{4}(m_{D^*}-m_D) \,.
\end{equation}
The relationship is not proven by experiments, but is compatible with several studies~\cite{PhysRevD.90.094007, PhysRevD.91.094502, PhysRevD.99.031501, Alexandrou:2017xwd}. Accordingly, the unknown mass of the $\Xi_{cc}^*$ can be determined. For the charm sector, from numerical analysis, $m_{D}=1868$ MeV, $m_{D^*}=2009$ MeV, $m_{\Sigma_c}=2454$ MeV, $m_{\Sigma^*_c}=2518$ MeV, $m_{\Xi_{cc}}=3570$ MeV,  and $m_{\Xi^*_{cc}}=3676$ MeV~\cite{PhysRevD.110.030001}.

The properties of a resonance, such as its mass and decay width, are influenced by its definition. This implies that any change in the definition of a resonance also affects its properties. The Breit-Wigner distribution formula is the most commonly used and accepted method for describing resonance and is expressed as follows:
\begin{equation}  \label{eq:9}
    d^{BW}(E)=\frac{\Gamma}{2\pi} \frac{1}{(E-M)^2+\frac{\Gamma ^2}{4}} \, ,
\end{equation} 
where $M$ is the mass, and $\Gamma$ is the decay width of the resonance. The Breit-Wigner formula is mainly used to describe unstable states. While it is suitable for most resonances, particularly fundamental particles like the $Z^0$ and $W^{\pm}$ bosons, it has limitations regarding near-threshold resonances and complex structures. For instance, the threshold effect can obscure the resonance region, leading to deviations from unitarity and analyticity. Additionally, the branching ratios derived from this parameterization for near-threshold states may not accurately represent actual decay probabilities~\cite{PhysRevD.93.034030}. Ultimately, the Breit-Wigner formula may not be suitable for near-threshold states~\cite{Cleven:2011gp}. To better understand near-threshold exotic resonances, it is essential to develop methods that are more suitable and applicable over a broader range of scenarios.

A few years ago, a new formalism was proposed to address the threshold issue inherent in the Breit-Wigner distribution, and the modified function was called Sill distribution~\cite{Giacosa:2021mbz}:
\begin{equation}  \label{eq:10}
     d^{Sill}(E)= \frac{2E}{\pi} \frac{\sqrt{E^2-E^2_{th}}\tilde{\Gamma}}{(E^2-M^2)^2+(\sqrt{E^2-E^2_{th}}\tilde{\Gamma})^2}\theta(E-E_{th}) \, ,
\end{equation}
where $\tilde{\Gamma}$ is defined as $\tilde{\Gamma}=\Gamma M / \sqrt{M^2-E^2_{th}}$. Notably, the Sill distribution contains threshold effects; it can be applied to mesons and exotics because the inner content of the hadron becomes irrelevant under this formalism. For various ranges of resonances, the Sill function provides more consistent results compared with experiments than those obtained under the Breit-Wigner formalism~\cite{Giacosa:2021mbz}. In  addition, the applicability of the Sill approach has been assessed. Compared with Breit-Wigner, Sill should be preferred when the $\Gamma  / \Delta m $ ratio is greater than $1/3$. Here, $\Delta m$ represents the distance between the state and threshold, and $\Gamma$ represents the decay width of the state. This criterion also applies to the $P_{c\bar{c}(s)}$ resonances studied herein.   

The LHCb Collaboration recently confirmed the existence of a signal for a particle called $P_{c\bar{c}s}(4338)$ from the $J/ \Psi \Lambda$ mass spectrum in the $B^-\rightarrow J/ \Psi \Lambda p$ process~\cite{LHCb:2022ogu}, showing that the particle has $J=1/2$. The mass of  $P_{cs}(4338)$  is close to the $\Xi_c \bar{D}$ threshold, but it exceeds the threshold. Therefore, its direct identification as the $\Xi_c \bar{D}$ molecular state was ruled out. Furthermore,  the shape of this resonance may potentially deviate from the conventional Breit-Wigner distribution if it strongly interacts with the threshold~\cite{Meng:2022wgl}, underscoring the need for further investigation. In addition to the $P_{c\bar{c}s}(4338)$ state, $P_{c\bar{c}s}(4459)$ has also been observed at the LHCb~\cite{LHCb:2020jpq}. This state may be considered a $\Xi_c\bar{D}^*(\frac{1}{2}/ \frac{3}{2})$ hadronic molecule because it has a small decay width and its mass is approximately $19$ MeV lower than the $\Xi_c\bar{D}^*$ threshold. 

The assessment of $P_{c\bar{c}s}(4338)$ as a hadronic molecule using the Breit-Wigner resonance model is inconvenient because its mass is above the threshold, but the Sill model may be a promising alternative. Upon applying the Sill model to the $P_{c\bar{c}s}(4338)$ resonance, its mass is pushed below the threshold, as shown in Table~\ref{tab:table2}, thereby enabling its evaluation as a molecule.
\begin{table}[htbp]
	\caption{ \label{tab:table2} Mass and decay widths $(m,  \Gamma)$  obtained by the Breit-Wigner parametrization's~\cite{PhysRevD.110.030001} and the Sill approach for the $P_{c\bar{c}}(4312)$,  $P_{c\bar{c}}(4440)$, $P_{c\bar{c}}(4457)$, $P_{c\bar{c}s}(4338)$  and $P_{c\bar{c}s}(4459)$ resonances. The values are all in units of MeV.  }
	\footnotesize
	\begin{tabular}{lll} 
		\toprule
	 	&  Breit-Wigner   & $\hspace{0.4 cm} $ Sill \\  \midrule  \\[-2.0ex]
		$P_{c\bar{c}}(4312)$  & $(4311.9,  9.8)$    &  $(4313.1,  5.0)$   \\
		$P_{c\bar{c}}(4440)$  &  $(4440.3,  20.6)$   &  $(4442.7,  10.6)$   \\
		$P_{c\bar{c}}(4457)$ &  $(4457.3,  6.4)$   &  $(4458.3,  3.3)$   \\
		$P_{c\bar{c}s}(4338)$ &  $(4338.2,  7.0)$   &  $(4329.5,  9.4)$    \\
		$P_{c\bar{c}s}(4459)$ &  $(4458.8,  17.4)$   &  $(4460.8,  8.9)$  \\ 
		\bottomrule
	\end{tabular}
\end{table}

To search for bound states, the Lippmann-Schwinger equation is useful because the solution corresponds to the mass of the bound state. In momentum space, the Lippmann-Schwinger equation can be written as
\begin{equation}\label{eq:11}
    \phi(k)+\int \frac{d^3p}{(2\pi)^3}\langle k\vert V \vert p \rangle \frac{\phi(p)}{B+\frac{p^2}{2\mu}}=0 \,,
\end{equation}
where $\phi(k)$ is the vertex function, $B$ is the binding energy, and $\mu$ is the reduced mass. For the regularization of the contact range potential with the $f(x)$ regulator function,  the potential is given by
\begin{equation}  \label{eq:12}
  \langle p\vert V \vert p^{\prime} \rangle = C(\Lambda) f(\frac{p}{\Lambda})f(\frac{p^{\prime}}{\Lambda})  \, ,
\end{equation}
where $f(x)=e^{-x^2}$ is the Gaussian regulator function chosen to regularize the ultraviolet divergence. The $\Lambda$ cutoff dependency of the low-energy constant is undesirable, but the dependency is generally mild and thus refitted into the contact terms. The $\Lambda$ cutoff is varied from $1.0$ to $1.5$ GeV, and significant changes in masses are not observed; therefore, the cutoff value of $1.0$ GeV is chosen for numerical calculations. In this context, dependency is not a concern.

The symmetries discussed herein are considered to be exact at the limit where the mass of the heavy quark approaches infinity. However, because the mass of heavy quarks is finite, some violations can be expected. For HQSS, the deviation is estimated to be $\mathcal{O}(\Lambda_{QCD}/ m_Q)$. For a $\Lambda_{QCD} $ of $\sim200$ MeV, this leads to an estimated deviation of approximately $15\%$ in the charm sector. In the case of HADS, the deviation is given as $\mathcal{O}(\Lambda_{QCD}/ (m_Q v))$, where $v$ represents the velocity of the heavy quark pair. According to a previous estimation~\cite{PhysRevD.73.054003}, the value of $m_Q v$ is approximately $800$ MeV. This results in a $25\%$ uncertainty in the HADS predictions. When both heavy quark symmetries are simultaneously considered, the total uncertainty is assigned  $ \delta =\sqrt{\delta^2_{HQSS} + \delta^2_{HADS} }= 29\%$.

For the $\bar{D}^{(*)}\Sigma_c^{(*)}$ molecular system, seven equations correspond to seven possible s-wave states. The four observed $P_{c\bar{c}}$ states can be considered HQSS partners: $\bar{D} \Sigma_c$ as $P_{c\bar{c}}(4312)$, $\bar{D} \Sigma^*_c$ as $P_{c\bar{c}}(4380)$, and molecular $\bar{D}^* \Sigma_c$ as $P_{c\bar{c}}(4440)$ and $P_{c\bar{c}}(4457)$. However, there is a problem with the interpretation of $P_{c\bar{c}}(4380)$ as a molecule because its decay width is considerably large, $205$ MeV. Hence, only the other states are labeled  as hadronic molecules. 

To estimate the full spectrum of these molecular states, it is necessary to determine two unknown counterterms: \(C_a\) and \(C_b\). To determine these two unknown low-energy constants, firstly, the masses of the $P_{c\bar{c}}(4440)$ and $P_{c\bar{c}}(4457)$ states are used as inputs. The two possible alternatives for these states' spins are labeled as \textit{options} $1$ and $2$.  \textit{Option} $1$ states that $P_{c\bar{c}}(4440)$ is $J^P=\frac{1}{2}^-$ and $P_{c\bar{c}}(4457)$ is $J^P=\frac{3}{2}^-$, whereas \textit{option} $2$ states the opposite, Table \ref{tab:table3}. In addition, some studies suggest that $P_{c\bar{c}}(4457)$ may not be exclusively suited to the $\bar{D}^*\Sigma_c$ molecule, but rather to a combination of spin $\frac{1}{2}$ and $\frac{3}{2}$ $\bar{D}^*\Sigma_c$ molecules~\cite{Xu:2020gjl}, or a mix of $\bar{D}^*\Sigma_c$ and $\Lambda_c (2595)\bar{D}$ states~\cite{Burns:2021jlu} or others~\cite{PhysRevD.100.014022,Chen:2019bip,PhysRevLett.122.242001,Chen:2019asm,PhysRevD.108.L031503, Wang:2022oof,PhysRevD.103.034003,Yamaguchi:2019seo, PhysRevD.103.054004,PhysRevD.102.114020,Uchino:2015uha,Karliner_2015}. At this point, another alternative is required to strengthen the analysis further. The mass of the $P_{c\bar{c}}(4312)$ state has not previously been used as an input in the literature; therefore, we incorporate the $P_{c\bar{c}}(4312)$ state along with the alternative spin states of \(P_{c\bar{c}}(4440)\) as \textit{options} $3$ and $4$. Specifically, \(P_{c\bar{c}}(4440)\) with \(J^P = \frac{1}{2}^-\) and  \(P_{c\bar{c}}(4440)\) with \(J^P = \frac{3}{2}^-\) are referred to as \textit{options} $3$ and $4$, respectively, in relation to the \(P_{c\bar{c}}(4312)\) state (Table \ref{tab:table3}). While \(P_c(4457)\) could also serve as a second input, we instead choose other option for the purposes of this study.

 \begin{table}[htbp]
	\caption{\label{tab:table3}  Input options employed in the numerical calculations}
	\footnotesize 	
	\begin{tabular}{ccc}
		\toprule 
		Option  & Input-$1$& Input-$2$   \\  \hline  \\[-2.0ex]   
		$1$ & $P_{c\bar{c}}(4440) \frac{1}{2}^-$ &  $P_{c\bar{c}}(4457) \frac{3}{2}^-$  \\[1.0ex] 
		$2$ & $P_{c\bar{c}}(4440) \frac{3}{2}^-$ &  $P_{c\bar{c}}(4457) \frac{1}{2}^-$  \\[1.0ex] 
		$3 $ & $P_{c\bar{c}}(4312) \frac{1}{2}^-$ &  $P_{c\bar{c}}(4440) \frac{1}{2}^-$  \\[1.0ex]  
		$4$ & $P_{c\bar{c}}(4312) \frac{1}{2}^-$ &  $P_{c\bar{c}}(4440) \frac{3}{2}^-$  \\[1.0ex]  
		\bottomrule
	\end{tabular}
\end{table}

With the relation between $P_{c\bar{c}}$ and $P_{c\bar{c}s}$, even though a direct relationship between the pentaquark states has not been observed, these states can have the same potential dependence (Table~\ref{tab:table1}), within the generalized flavor-spin symmetry. Thus, we can predict possible $\bar{D} \Xi_c$ and  $\bar{D}^* \Xi_c$ molecular states for $P_{c\bar{c}s}$  that share the same low-energy constants. To establish this association, the potentials of $P_{c\bar{c}}(4440)$ and $P_{c\bar{c}}(4457)$ are assumed to be $\frac{1}{2}$ and $\frac{3}{2}$, respectively.

\section{Results and Discussions}\label{sec:3}

As shown in Table~\ref{tab:table4}, all states are classified as bound states for the first two options, even when considering the uncertainty of the HQSS. All  states are also located at or near the threshold, suggesting that they could be probed in future experiments. Up to this point, the results are consistent within the considered uncertainties because the mass of $P_{c\bar{c}}(4312)$ is successfully obtained within the uncertainty limits, and for \textit{option} 1, the mass of $P_{c\bar{c}}(4380)$ is estimated with sufficient accuracy. In addition, the expectation that a high-spin state has a heavy mass is observed for $\bar{D}^* \Sigma_c^*$ states, with considerable accuracy. However, detecting the three $\bar{D}^{*}\Sigma_{c}^{*}$ states poses a challenge owing to their low production rates in terms of  $\Lambda_b$ decay, which has been previously observed~\cite{PhysRevLett.124.072001}. The $\frac{5}{2}^-$ $\bar{D}^*\Sigma_c^*$ state is also unlikely to be observed in the LHCb experiment, given that it does not couple to the $J/ \Psi p$ in the s-wave~\cite{PhysRevD.100.014021}. Conversely, for \textit{options} 3 and 4, all states are also bound, except for the $\frac{1}{2}^-$  $\bar{D}^*\Sigma_c^*$  state in \textit{option} 4, with a threshold value of $4527$ MeV.  In particular, the mass of $P_{c\bar{c}}(4457)$ in \textit{option} 3 is similar to the experimental result. As shown in Eq.~\ref{eq:4}, $C_a$  does not depend on the spin, whereas $C_b$ represents the spin-spin interaction. The effects of spin-spin interactions are seen in the $\bar{D}^{(*)}\Sigma^{(*)}$ states.
\begin{table}[htbp]
	\caption{\label{tab:table4} The leading order contact range potential derived from HQSS for the heavy meson and heavy baryon system depends on the linear combination of two coupling constants, $C_a$ and $C_b$. These coupling constants are determined by reproducing masses of the $P_{c\bar{c}}(4312)$, $P_{c\bar{c}}(4440)$,  and $P_{c\bar{c}}(4457)$ with each possible option.  For the options we find that the couplings are  $C_a=-0.7909$ fm$^2$ and  $C_b=0.1048$ fm$^2$;  $C_a=-0.8608$ fm$^2$ and  $C_b=0.1048$ fm$^2$;   $C_a=-0.8145$ fm$^2$ and  $C_b=0.0871$ fm$^2$;  $C_a=-0.8145$ fm$^2$ and  $C_b=-0.1742$ fm$^2$; respectively. The errors come from the uncertainty of HQSS. Mass results are given in MeV. } 
\footnotesize
\begin{tabular}{lccllll}   
	\toprule
	Molecule  & $J^P$  & V & \textit{Option} 1 & \textit{Option} 2 & \textit{Option} 3 & \textit{Option} 4   \\[1ex] \midrule \\[-2.0ex]
	$\bar{D}\Sigma_c$ &$C_a$   & $\frac{1}{2}^-$ &  $4313^{-10}_{+6}$  &  $4308^{-12}_{+9}$ &   $P_{c\bar{c}}(4312)$  &  $P_{c\bar{c}}(4312)$  \\[1ex] 
	$\bar{D}\Sigma_c^*$ &  $C_a$ & $\frac{3}{2}^- $ &  $ 4378^{-10}_{+6}$   &  $4372^{-13}_{+9} $     &  $4376^{-10}_{+8}$   &    $4376^{-10}_{+8}$   \\[1ex]  
	$\bar{D}^*\Sigma_c$  &  $C_a-\frac{4}{3}C_b$ & $\frac{1}{2}^-$ & $ P_{c\bar{c}}(4440)$  &  $ P_{c\bar{c}}(4457)$    & $ P_{c\bar{c}}(4440)$  &  $4462^{-2}_{+1}$   \\[1ex] 
	$\bar{D}^*\Sigma_c$  & $C_a+\frac{2}{3}C_b$& $\frac{3}{2}^- $&  $ P_{c\bar{c}}(4457)$ &   $ P_{c\bar{c}}(4440)$   & $4456^{-9}_{+6}$   & $ P_{c\bar{c}}(4440)$   \\[1ex]  
	$\bar{D}^*\Sigma_c^*$  &$C_a-\frac{5}{3}C_b$&  $\frac{1}{2}^-$&  $4501^{-17}_{+13} $ &  $4524^{-7}_{+3}$   & $4501^{-16 }_{+14}$  &  $4527 $   \\[1ex] 
	$\bar{D}^*\Sigma_c^*$ & $C_a-\frac{2}{3}C_b$ &$\frac{3}{2}^-$ & $4511^{-13}_{+10}$  &  $4517^{-11}_{+6}$ &  $4510^{-13}_{+10}$  &  $4523^{-7}_{+4}$   \\ [1ex]
	$\bar{D}^*\Sigma_c^*$  & $C_a+C_b$  & $\frac{5}{2}^-$& $4524^{-7}_{+3}$  &  $4501^{-17}_{+13}$ &  $4522^{-8}_{+5}$  &  $4498^{-17}_{+15}$   \\  
	\bottomrule
\end{tabular}
\end{table}
Assuming that $P_{c\bar{c}}(4440)$ and $P_{c\bar{c}}(4457)$ are hadronic molecules, ten HADS partners of the $\Xi_{cc}^{(*)} \Sigma_c^{(*)}$ system are studied herein (Table~\ref{tab:table5}). Considering all the alternative options, all  molecule states are bound under the total uncertainty of the HQSS and HADS, consistent with a previously study showing that all HADS partners are bound~\cite{Pan:2020xek}. However, studies have shown various outcomes for each state. First, our results regarding the existence of four of  the ten states, namely $\Xi_{cc}\Sigma_c$ with $0^+$ and $1^+$ and $\Xi_{cc}\Sigma_c^*$ with $1^+$ and $2^+$, are theoretically supported by previous work~\cite{PhysRevD.97.114011}. Furthermore, a lattice QCD study on heavy dibaryons concluded that the $1^+$ $\Xi_{cc}\Sigma_c$ state has a binding energy of  $8\pm 17$ MeV~\cite{PhysRevLett.123.162003}. One study tried to determine s-wave triply charmed dibaryons in a model-independent way, concluding that the $1^+$ $\Xi_{cc}\Sigma_c$ state has a binding energy of $15-30$ MeV~\cite{PhysRevD.102.011504}. Another study~\cite{Pan:2019skd} related the mass splitting of $P_{c\bar{c}}(4440)$ and $P_{c\bar{c}}(4457)$ with $0^+$ $\Xi_{cc}\Sigma_c$ and $1^+$ $\Xi_{cc}\Sigma_c$, reaching the same conclusion about the $1^+$ $\Xi_{cc}\Sigma_c$ state, consistent with our findings and predictions made by other researchers. Moreover, the expectation that a higher spin state has a heavier mass can be seen for $\Xi_{cc}^* \Sigma_c^*$ states.  \textit{Options} $1$ and $3$ met this expectation.  
\begin{table}[htbp]
	\caption{ \label{tab:table5} The leading order contact range potential derived from HQSS for the doubly heavy baryon and heavy baryon system depends on the linear combination of two coupling constants, $C_a$ and $C_b$. From HADS, the relationship of coupling constants $C_a$ and $C_b$ are the same as Table~\ref{tab:table4}. The errors come from the total uncertainty of HADS and HQSS. Mass results are given in MeV.}
	\begin{tabular}{lclccccc}  
		\toprule
		Molecule & $J^P$ & V &  \textit{Option} 1 & \textit{Option} 2  & \textit{Option} 3 &   \textit{Option} 4 & Threshold   \\[1ex] \midrule  \\[-2.0ex]
		$\Xi_{cc}\Sigma_c$ & $0^+$ &  $C_a+\frac{2}{3}C_b$ & $6005^{-25}_{+17}  $  & $ 5980^{-38}_{+31}  $  & $6002^{-14}_{+11} $  &  $5980^{-19}_{+18}  $ & \multirow{2}{*}{$6024$}  \\[1.0ex]  
		$\Xi_{cc}\Sigma_c$ & $1^+$ &  $C_a-\frac{2}{9}C_b$ & $ 5995^{-34}_{+23}  $  & $ 5992^{-32}_{+25} $   & $5993^{-16}_{+14}  $  &  $ 5999^{-14}_{+12} $ &   \\[1ex]  
		$\Xi_{cc}\Sigma_c^*$ & $1^+$ &  $C_a+\frac{5}{9}C_b$ & $ 6068^{-26}_{+18}  $  & $ 6046^{-39}_{+30} $  &  $6064^{-14}_{+12}  $ & $ 6047^{-19}_{+16} $ &   \multirow{2}{*}{$6089$} \\[1ex] 
		$\Xi_{cc}\Sigma_c^*$ & $2^+$ &  $C_a-\frac{1}{3}C_b$ & $ 6057^{-32}_{+24}  $ & $ 6057^{-32}_{+24} $ &  $ 6055^{-16}_{+15} $  &   $  6065^{-13}_{+12} $ & \\ [1ex]  
		$\Xi_{cc}^*\Sigma_c$ & $1^+$ &  $C_a-\frac{10}{9}C_b$ & $ 6088^{-37}_{+30}  $  & $ 6108^{-27}_{+18} $ &  $6088^{-19}_{+17}  $ &  $6120^{-9}_{+6}  $  & \multirow{2}{*}{$6130$}  \\[1ex] 
		$\Xi_{cc}^*\Sigma_c$ & $2^+$ &  $C_a+\frac{2}{3}C_b$ & $ 6111^{-26}_{+16}  $  & $ 6085^{-39}_{+31} $ &  $  6107^{-16} $  &  $ 6085^{-19}_{+18} $ &  \\[1.0ex] 
		$\Xi_{cc}^*\Sigma_c^*$ & $0^+$ &  $C_a-\frac{5}{3}C_b$ & $  6144^{-42}_{+34}  $  & $ 6178^{-23}_{+15} $  & $ 6145^{-21}_{+19}  $  & $6191^{-5}_{+3}  $ &  \multirow{4}{*}{$6194$}  \\[1.2ex] 
		$\Xi_{cc}^*\Sigma_c^*$ & $1^+$ &  $C_a-\frac{11}{9}C_b$ & $ 6150^{-38}_{+31}  $  & $ 6173^{-26}_{+18} $  &  $6150^{-19}_{+17}  $   & $ 6186^{-9}_{+6} $ &   \\[1ex] 
		$\Xi_{cc}^*\Sigma_c^*$ & $2^+$ &  $C_a-\frac{1}{3}C_b$ & $ 6162^{-32}_{+25}  $  & $ 6162^{-32}_{+25} $  &  $ 6160^{-17}_{+15}  $  &  $ 6171^{-14}_{+11}  $ & \\[1ex] 
		$\Xi_{cc}^*\Sigma_c^*$ & $3^+$ &  $C_a+C_b$ & $ 6178^{-23}_{+15} $  & $ 6144^{-42}_{+34}  $ &  $ 6174^{-13}_{+10} $   &  $6141^{-22}_{+20}  $ &    \\[1ex]
		\bottomrule
	\end{tabular}
\end{table}
Evaluation of the $ P_{c \bar{c}s}(4338)$ state as a molecule is controversial because its Breit-Wigner mass is above the $ \bar{D} \Xi_c$ threshold. The negative binding energy problem has generally been ignored when examining a hadronic molecule, but when the Sill parametrization is applied, the problem is solved. This pushed the mass below the related threshold (Table~\ref{tab:table2}), making it suitable for to evaluating the $ \bar{D} \Xi_c$ molecule. The Sill mass value of the $P_{c \bar{c}s}(4338)$ is supported by previous works~\cite{Chen:2021cfl, Chen:2022wkh, Giachino:2022pws} and agrees with all options derived from flavor-spin symmetry, except \textit{option} $2$, as shown with uncertainties in Table~\ref{tab:table6}.

\begin{table*}[htbp]
	\caption{\label{tab:table6}  Predicted mass values of $P_{c\bar{c}s}(4338)$ and $P_{c\bar{c}s}(4459)$. Mass results are given in MeV.}
	\footnotesize
	\begin{tabular}{@{}c|ccccccc}  
		\toprule
		Molecule &   \textit{Option} 1  &   \textit{Option} 2&  \textit{Option} 3 &   \textit{Option} 4 &  Ref.~\cite{Chen:2021cfl} & Ref.~\cite{Chen:2022wkh}  & Ref.~\cite{Giachino:2022pws} \\[1ex]  \midrule  \\[-2.0ex]
		$P_{c\bar{c}s}(4338)$ &   $4329^{-9}_{+7}$ &$4324^{-12}_{+9}$ &$4328^{-5}_{+7}$ & $4328^{-11}_{+7}$ & $4327.7$ &  $ 4328.5$  &  $4329.1 $\\[1ex]   
		$P_{c\bar{c}s}(4459)$  &$4469^{-11}_{+7}$ & $ 4463^{-13}_{+10}$ &  $4467^{-12}_{+8}$ & $4465^{-12}_{+10}$  & $4468.1 $ & $4468.3$  & $ 4469.2$ \\  
		\bottomrule
	\end{tabular}
\end{table*}

When the available data increased, $P_c(4450)$ consisted of two peaks. Some studies have suggested that the same situation may be valid for $P_{c \bar{c}s}(4459)$~\cite{Wang:2022mxy, Karliner:2022erb}. $ P_{c\bar{c}s}(4459)$  may hide behind the $P_{c \bar{c}s}(4455)$ and $P_{c \bar{c}s}(4468)$  states as $\Xi_c\bar{D}^*$ molecules. The masses obtained in our results for $P_{c\bar{c}s}(4338)$ and $P_{c\bar{c}s}(4468)$ are consistent with the results of other studies, suggesting that $P_{c\bar{c}s}(4459)$ can be explained by a two-peak interpretation, as shown in Table~\ref{tab:table6} ~\cite{Chen:2021cfl,Giachino:2022pws,Ke:2023nra}. In this context, it is important to note that the results obtained from \textit{options} $2$ and $4$ are more distinct. These two options clearly show tension with other options.

\section{Summary} \label{sec:4}

In this study, we apply heavy quark symmetries to determine the spins of the $P_{c\bar{c}}(4440)$ and $P_{c\bar{c}}(4457)$ states. Initially, $P_{c\bar{c}}(4440)$ and $P_{c\bar{c}}(4457)$ are considered  $\bar{D}^*\Sigma_c$ hadronic molecules; hence, we can classify them as HQSS partners. Second, the first two possible assignments are presented to clarify the ambiguity regarding their spin states. All $\bar{D}^{(*)}\Sigma_c^{(*)}$ states are found to be bound under the first two options, suggesting that they may be accessible in future experiments. Then, the $P_{c\bar{c}}(4312)$ state is included as a molecule, and the spin options of $P_{c\bar{c}}(4312)$ with $P_{c\bar{c}}(4440)$ are used to specify the possible spin state for the last two options. Among these four options, the mass of $P_{c\bar{c}}(4380)$ is significantly close to that predicted by \textit{option} 1$(4378  \text{ MeV})$, and the mass of the $ P_{c\bar{c}}(4457)$ aligns with \textit{option} 3$(4456  \text{ MeV})$. Moreover, the prediction regarding $\bar{D}^{(*)}\Sigma_c^{(*)}$ states suggests that $P_{c\bar{c}}(4440)$ has a spin of $\frac{1}{2}^-$ and $P_{c\bar{c}}(4457)$  has a spin of  $\frac{3}{2}^-$. Next, HADS partners are consulted to examine the HQSS findings. The $\Xi_{cc}^{*}\Sigma_c^{*}$ HADS partners corresponding to \textit{options} 1 and 3 show a similar behavior where higher mass states have higher spin states. Finally, the generalized spin-flavor symmetry is applied. The similarity between $P_{c\bar{c}}$ and $P_{c\bar{c}s}$ implies that the $ \bar{D}^{(*)}\Sigma_c^{(*)}$ and $\bar{D}^{(*)}\Xi_c^{(*,\prime)}$ states share the same potential. When the updated resonance parameterization is applied to $P_{c\bar{c}s}(4338)$, the consistency of the results improves, allowing for a more reliable application of the same potentials. However, even when this symmetry is applied, the results provide limited constraints on the spin estimations of the $ P_{cc} $ states because they conflict with the observed mass values of $P_{c\bar{c}s}$. Nonetheless, the obtained masses  help eliminate one possible spin state while agreeing well, especially in \textit{options} $1$ and $3$, with many results reported in the literature. 

This study is based on several assumptions that could influence the final results. 
First, a molecular interpretation of the $P_{c\bar{c}}$ states is adopted. 
Second, our analysis focuses on contact-range interactions, while possible coupled-channel effects and long-range interactions are neglected. Additionally, since HQSS and HADS are approximate, their potential violations introduce inherent uncertainties. 
Therefore, the obtained spin assignments should be interpreted with caution within the scope of these approximations.
However, the results also indicate that $P_c(4450)$ contains two other states. Taken together, $P_c(4459)$ appears to share similar characteristics with $P_c(4450)$ and likely consists of two states. In conclusion, our results favor the assignment of $J^P = \frac{1}{2}^-$ for the $P_{c\bar{c}}(4440)$ state and $J^P = \frac{3}{2}^-$ for the $P_{c\bar{c}}(4457)$ state, although uncertainties remain. The discovery of any of the $\Xi_{cc}^{(*)}\Sigma_c^{(*)}$, $P_{c\bar{c}}$, and $P_{c\bar{c}s}$ states significantly strengthens our assumptions. Further studies on heavy exotics and  LHC runs are needed to make  more precise estimations of hadronic molecules.

\backmatter
\bmhead{Conflict of interest}
The author received no financial support for the research, authorship, and/or publication of this article. The author has no conflicts of interest to disclose. The author declares that no data had been used nor generated supporting the findings of this study.


\smallskip

\end{document}